# Thermodynamic signatures of a potential Fulde-Ferrell-Larkin Ovchinnikov state in the isotropic superconductor $Ti_4Ir_2O$

Jiaqi Hu[1], Yat Hei Ng[1], Omargeldi Atanov[1], Bin-Bin Ruan[2], Zhi-An Ren[2,3], and Rolf Lortz[1,*]

[1]Department of Physics, The Hong Kong University of Science & Technology, Clear Water Bay, Kowloon, Hong Kong, China
[2]Institute of Physics and Beijing National Laboratory for Condensed Matter Physics, Chinese Academy of Sciences, Beijing 100190, China
[3]School of Physical Sciences, University of Chinese Academy of Sciences, Beijing 100049, China

[*]Correspondence to Rolf Lortz (lortz@ust.hk)

## Abstract

This study presents an investigation of the isotropic superconductor $Ti_4Ir_2O$, focusing on its magnetic field vs. temperature phase diagram near and above the Pauli limit for superconductivity. The data exhibits characteristic features that align with the formation of a Fulde-Ferrell-Larkin-Ovchinnikov (FFLO) state, including a deviation of the upper critical field line from the standard WHH model upon approaching the Pauli limiting field and a phase transition line separating the ordinary low-field superconducting phase from an unusual high field superconducting state above the Pauli limit. We discuss why the upper critical field line in this isotropic superconductor can approach the Pauli limit, necessitating a particularly high orbital limit for superconductivity. The research also identifies peculiar features in the specific heat data set taken at 0.6 K, which may be an example of the segmentation of the FFLO phase with multiple **q** phases. The findings provide an explanation for the violation of the Pauli limit in this superconductor, adding a fully isotropic superconductor to the list of potential FFLO superconductors and providing new insights for contemporary theories of the FFLO state.

## Introduction

The Fulde-Ferrell-Larkin-Ovchinnikov (FFLO) state [1,2] is a distinct superconducting phase observed in spin singlet superconductors when subjected to very high magnetic fields beyond the Pauli limiting magnetic field for superconductivity [3,4]. This FFLO state, which holds significant technological relevance in high magnetic field applications, has been observed in only a few superconducting materials. These include the layered heavy-fermion superconductor $CeCoIn_5$, where it coexists with a spin density wave, forming the prominent Q-phase [5-7], layered organic superconductors [8-18] and additional instances of this phase have also been reported in iron-based superconductors such as $KFe_2As_2$ [19] and FeSe [20], and recently in the transition metal dichalcogenide superconductors $NbS_2$ [21] and $NbSe_2$ [22].

The Pauli limiting magnetic field represents the theoretical maximum limit for superconductivity. At this critical field, the Zeeman splitting energy between two electrons with opposite spins - which form the Cooper pair - reaches a level that abruptly prevents further pair formation. This phenomenon triggers a sudden, discontinuous first-order phase transition, causing the system to return to its normal state [6,9,10]. However, observing this phenomenon is rare, as most spin-singlet superconductors do not reach this Pauli limit. In fact, the upper critical field for the majority of type-II superconductors typically occurs at much lower applied magnetic fields. This deviation is primarily due to the orbital effect, where the superconducting screening currents reach a threshold that disrupts the pairing, leading to a continuous second-order transition into the normal state [23].

The majority of known FFLO superconductors possess highly anisotropic layered structures [5-22]. When a magnetic field is applied parallel to these layers, the orbital limit for superconductivity is significantly increased, potentially surpassing the Pauli limit. In the magnetic field versus temperature phase diagram, the upper critical field transition line typically exhibits a steep initial slope, followed by a plateau at the Pauli limiting field due to the abrupt suppression caused by the Zeeman effect [6,9,10,19,21]. However, the formation of the FFLO state offers a solution for the superconductor to maintain its superconducting state. This is achieved by forming Cooper pairs with a finite center-of-mass momentum, resulting in a kind of pair density wave state [24]. In this state, the amplitude of the superconducting order parameter undergoes spatial modulation, allowing the superconductor to exist in magnetic fields that extend beyond the Pauli limit.

Recently a possible violation of the Pauli limit has been reported in the isotropic superconductor Ti$_4$Ir$_2$O, which possesses a cubic structure of the space group $Fd\bar{3}m$. The superconductivity in this material has been observed to persist up to approximately 1.5 times the weak-coupling Pauli limiting field [25,26]. While no definitive evidence of an FFLO state has been reported thus far, our article presents high-resolution data of the magnetization and the specific heat that suggest the possible formation of the FFLO state beyond the Pauli limit based on bulk thermodynamic probes. The primary experimental indicators for the formation of an FFLO state typically emerge from thermodynamic probes. These probes reveal a characteristic upturn in the upper critical field line within the low-temperature, high-magnetic-field regime of the magnetic phase diagram. Additionally, they enable the identification of a phase transition line that separates the ordinary superconducting state from the FFLO state [5-22]. In this article we present these key indicators: Our observations reveal that the upper critical field initially exhibits a slight deviation from the Werthamer-Helfand-Hohenberg (WHH) model near the Pauli limit at 8 T, displaying a weak saturating trend. However, the slope increases slightly again, with superconductivity persisting beyond 14 T. This behavior aligns with the anticipated upturn of the upper critical field line, indicative of the formation of an FFLO phase. The upturn in this isotropic superconductor is, however, much weaker than in most known layered FFLO superconductors [8-22]. Notably, the specific heat exhibits a distinct phase transition within the superconducting state above the Pauli limit, accompanied by evident hysteresis during field sweeps. This along with the slope change of the upper critical field line at the Pauli limit suggests that an FFLO state may be realized in Ti$_4$Ir$_2$O in a fully isotropic superconductor.

# Results

**Magnetization experiments**

Fig. 1 presents the magnetization measurements obtained using the Faraday balance. It is important to note that the data is presented in arbitrary units due to the challenges in determining the precise absolute magnitude of the force acting on the cantilever during the measurement. Our focus is primarily on the shape of the magnetization curves, rendering the absolute value irrelevant. For clarity, we have subtracted a perfectly linear and temperature-independent normal state background, which reaches a magnitude of less than 10% of the maximum superconducting signal at 14 T, from all data.

The magnetization curves exhibit a pronounced hysteresis upon sweeping the field up and down. This is attributed to the opposite signs of the induced screening currents, which reverse the direction of the force on the cantilever. At temperatures above 3 K, the magnetization curves display the characteristic continuous second-order transition at $H_{c2}$, where the superconducting contribution to the magnetization vanishes in the form of a kink. However, at lower temperatures, where the $H_{c2}$ values begin to exceed 7 T, a broad peak develops below the upper critical field, followed by an abrupt drop in magnetization to zero at $H_{c2}$. This transformation of the $H_{c2}$ transition into a step-like anomaly has also been reported in other superconductors that approach or exceed the Pauli limit [10,19,21]. The $H_{c2}$ values have been determined at different temperatures at the highest field below which the magnetization deviates from the linear normal state background (as shown in detail for $T = 0.33$ K and 0.8 K in Fig. 1b&c), which appears as a kink-like onset of the Meissner diamagnetism, above the onset of irreversibility due to flux pinning marked as $H_{irr}$. The combination of relatively sharp features and high resolution of this technique enables us to precisely determine the upper critical field line for subsequent inclusion in a phase diagram. Notably, the kinks define the $H_{c2}$ values with error bars smaller than 0.1 T.

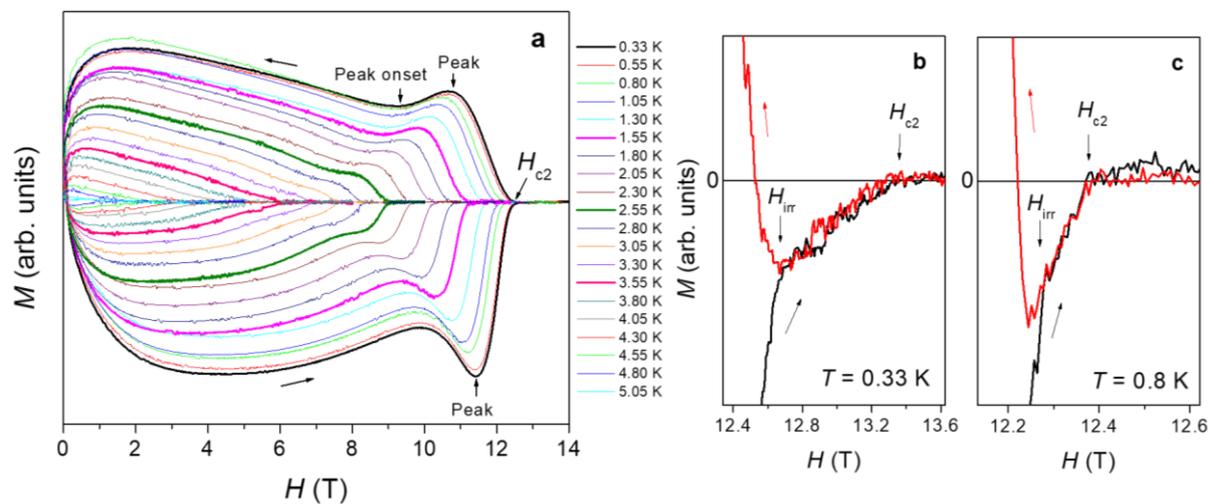

**Fig. 1. a** Magnetization data of Ti$_4$Ir$_2$O measured with a Faraday balance at different fixed temperatures as a function of increasing and decreasing magnetic field as marked by the arrow. A small linear normal state contribution has been removed. An additional peak like anomaly is marked at the field value where the peak is centered and at the lower onset. The upper critical field $H_{c2}$ is defined by the deviation from the normal state background as indicated for the 0.33 K data. Panel (**b**) and (**c**) show enlarged views onto the data near the upper critical field $H_{c2}$ at 0.33 K and 0.8 K, respectively. The field below which irreversibility due to flux pinning occurs is indicated as $H_{irr}$.

### Specific heat experiments

In our AC calorimetry measurements, the sample temperature is sinusoidally modulated with a small amplitude not exceeding 1 mK. This allows us to conduct measurements both as a function of a varying base temperature, and at a fixed base temperature as a function of the magnetic field. Note, that in both cases the measurements probe the thermodynamic response of the sample with respect to a small temperature change. This field sweep method is particularly suitable here to look for magnetic field induced phase transition at low temperatures.

To quickly characterize the sample, in Fig. 2 we show zero field specific heat data, which agree perfectly with data in Ref. 26, showing the superconducting transition at $T_c = 5.1$ K. We thus refer to Ref 26 where relevant superconducting parameters have been derived, which agreed with a strong coupling superconductivity. The electronic Sommerfeld constant and the specific heat jump are slightly larger than in Ref. 25, but we found that upon normalizing the electronic specific heat $C/T$ by the Sommerfeld constant and plotting it over a reduced temperature $T/T_c$, their data perfectly match ours. The data allows us to obtain an accurate estimation of the Pauli limit. We analyze it with the standard alpha model to obtain the ratio $2\Delta(0)/k_BT_c$. The data in the available temperature range can be reasonably well described by a single gap s-wave alpha model with $2\Delta(0)/k_BT_c = 3.8$. In Ref. 26 it was suggested that $Ti_4Rr_2O$ may be a multi-gap superconductor. While we cannot exclude the possibility that data at lower temperatures may be necessary to observe the contribution of the second gap, the larger gap, which determines the Pauli limit, provides us with sufficient information. Note that our gap value can be applied equally well to fit the specific heat data in Ref. 25. We use the value of $2\Delta(0)/k_BT_c = 3.8$ to calculate the Pauli limit following Clogston's approach $H_P = \Delta(0) / \sqrt{2}\mu_B = 8.2$ T [3].

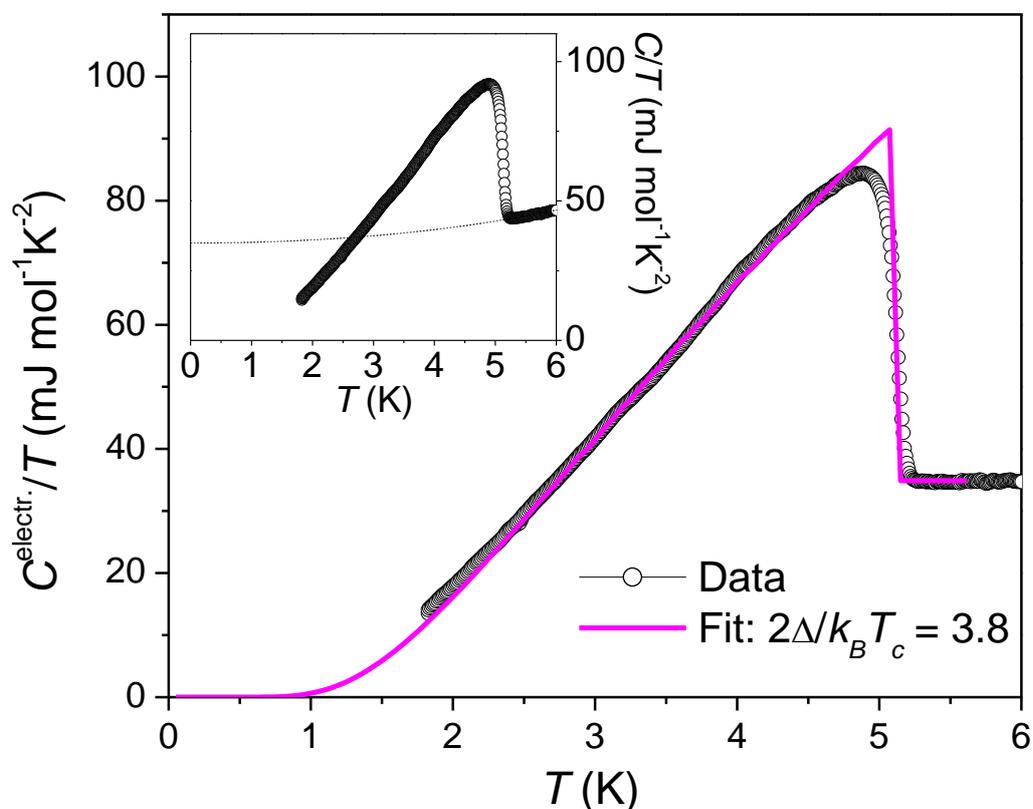

**Fig. 2.** Electronic contribution to the specific heat $C^{electr.}/T$ of $Ti_4Ir_2O$ in zero-field. The magenta line represents a fit with the alpha model with a superconducting gap of $2\Delta(0)/k_BT_c = 3.8$. The inset shows the total specific heat together with a fit to the normal state background.

Fig. 3a&b present the field dependent specific heat measurements at a few selected temperatures up to 1.5 K measured upon increasing (a) and decreasing field (b). Data at higher temperatures are shown separately in Fig. 3c. The primary jump-like anomaly represents the upper critical field transition, beyond which the normal state is restored. For the lowest temperature data we have marked the midpoint of this broadened jump as $H_{c2}$, which agrees perfectly with the $H_{c2}$ values obtained from the magnetization onset in Fig. 1. This makes sense as $C_p$ is a second derivative of the free energy for which the superconducting transition represents a jump like anomaly, which is typically broadened in high fields due to field induced

fluctuations [27]. The midpoint corresponds then to the field or temperature where the phase coherence is established, which coincides with the formation of a zero-resistance state. In contrast, magnetization is the first derivative of free energy, and the upper critical field transition is largely dominated by the decay of screening currents, which must vanish at the point where phase coherence is lost. As we will show later in the form of a phase diagram these two definitions of the superconducting transition in a field match very well.

Interestingly, an additional jump is observed in Fig. 3a slightly below the upper critical field transition, exhibiting a clear hysteresis upon sweeping the field up and down. This second transition, which we attribute to the transition between the ordinary low-field phase and the FFLO phase in high fields, is visible between 0.6 K and 1.5 K. In the data measured at 0.6 K and 0.8 K, two distinct additional transition anomalies are visible as marked by the arrows. These additional phase transformations are no longer visible at 2 K and above (Fig. 3c). Fig. 3d shows selected $C/T$ data measures as a function of temperature in constant fields of 9 and 9.5 T, where a similar kink like anomaly is visible below the main superconducting transition, as marked by the arrows.

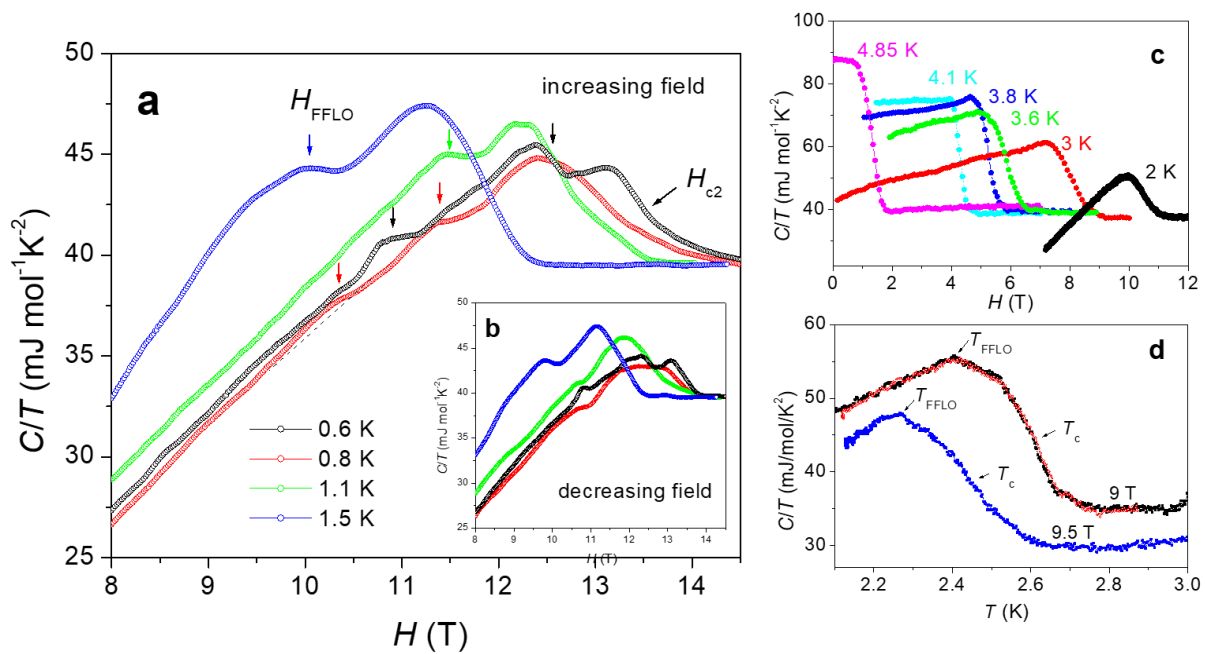

**Fig. 3.** High magnetic field specific heat $C/T$ of $Ti_4Ir_2O$ measured with an AC calorimetry technique. **a** $C/T$ measured at fixed temperatures as a function of increasing magnetic field at low temperatures up to 1.5 K where the potential FFLO phase occurs. The upper critical field transition at $H_{c2}$ is taken from the midpoint of the main broadened specific heat jump, while additional small anomalies are marked by vertical arrows as transitions into the high field FFLO phase. At the lowest temperatures two separate FFLO transitions become visible (see text for details). **b** Similar data as in (a) but measured upon decreasing field. **c** $C/T$ measured at higher temperatures where no more FFLO phase occurs. **d** $C/T$ measured in 9 T and 9.5 T (shifted by -5 mJ/mol/K$^2$ for clarity) across the FFLO transition at $T_{FFLO}$ (see arrows) as a function of temperature.

# Discussion

Fig. 4 presents the phase diagram of $Ti_4Ir_2O$, which plots magnetic field against temperature, compiled from all the collected data. The Pauli limit obtained from $H_p = \Delta(0) / \sqrt{2}\mu_B$ is represented by a dashed line at a constant field of 8.4 T. A dotted fit to the high-temperature upper critical field data using the Werthamer–Helfand–Hohenberg (WHH) model is also included. This model fits the data very well up to about 8 T. However, above 8 T, the $H_{c2}$ data, determined from the magnetization and specific heat deviates from the WHH model and remains below it in the high field range, which appears to correlate with our calculated estimate of $H_p$. Note that the only fitting parameters in the WHH model are the zero field $T_c$ and the zero temperature $H_{c2}$ values. The abrupt change in the slope of the $H_{c2}$ line is more distinctly illustrated by the jump-like anomaly observed in the temperature derivative of the $H_{c2}$ line. This anomaly is depicted in the inset of Fig. 4. It reveals that the slope of the $H_{c2}$ line undergoes a sudden reduction at this field, and subsequently continues to escalate in higher fields, approximately coinciding with the crossover to a more jump-like anomaly at $H_{c2}$ observed in the magnetization. It should be noted that the WHH model provides a significantly improved fit to our data compared to Ref. [26] where the fit was notably below the $H_{c2}$ data. This discrepancy is likely due to the much sharper definition of $H_{c2}$ in our magnetization data (see Fig. 1 b & c). The application of an alternative model to describe the temperature dependence of the $H_{c2}$ line would not yield superior results [26]. For instance, the Ginzburg-Landau relation would result in an even more linear relationship, while a multigap model could potentially induce an upturn. In contrast, our data exhibits a relatively abrupt downturn deviation above 8 T, which can be plausibly attributed to the paramagnetic pairbreaking effect in proximity to the Pauli limit.

Below 2.3 K, the $H_{c2}$ line exceeds the Pauli limit and continues to increase, steeper than the WHH line, until the slope gradually begins to decrease below 1.5 K. This behavior appears distinct from other known FFLO superconductors realized in layered 2D materials, where the $H_{c2}$ line first tends to saturate at the Pauli limit but then shows a pronounced upturn when the FFLO phase is formed. However, the deviation from the WHH model and the subsequent increase indicates a similar, albeit weaker effect. The deviation from the WHH model is attributed to the onset of Zeeman pair breaking effects when the Pauli limit is approached. Meanwhile, the linear slope, which becomes steeper than the WHH line in a range above the Pauli limit, illustrates the strengthening of the superconducting state with respect to the paramagnetic effect once the FFLO phase is established.

The most compelling evidence for the formation of the FFLO state is the relatively sharp phase transition anomalies in the specific heat. These anomalies indicate the existence of an unusual superconducting phase that exists in a field range above the Pauli limit. The anomaly reduces the specific heat, which distinguishes it from a vortex melting transition, where the specific heat increases in the liquid phase. This behavior agrees with that a part of the condensate is lost to the normal state due to the order parameter modulation of the FFLO state. We thus attribute these phase transitions to the transition separating the high-field FFLO phase from the ordinary superconducting phase in the low-field range. This is because it shares many similarities with the FFLO transition lines in other layered FFLO superconductors [5-22]. The transition exhibits clear hysteretic behavior, similar to that observed at the FFLO transition in organic superconductors [10]. Sample inhomogeneity could potentially cause two separate superconducting transitions in zero field and two distinct upper critical field transitions without significant hysteresis, as the upper critical field transition of a superconductor typically does not display hysteresis as observed here. Our zero-field specific heat measurements reveal only a single superconducting transition of reasonable sharpness, with no indication of an impurity phase with a different $T_c$. Therefore, these small transitions cannot be attributed to an impurity superconducting phase. Other explanations for the unusual phase transition above the Pauli limit within the superconducting state are more exotic than the FFLO state, such as spin triplet superconductivity. This scenario is less likely than an FFLO state and does not account for why our additional phase transition line begins approximately at the Pauli limit. Please note that no distinct anomalies are observed at $H_P$ in the magnetization measurements. However, a pronounced peak effect is evident in this field range, likely due to the weakening of the vortex lattice as the FFLO state forms. The resulting strong screening currents obscure any minor thermodynamic signatures of the phase transition.

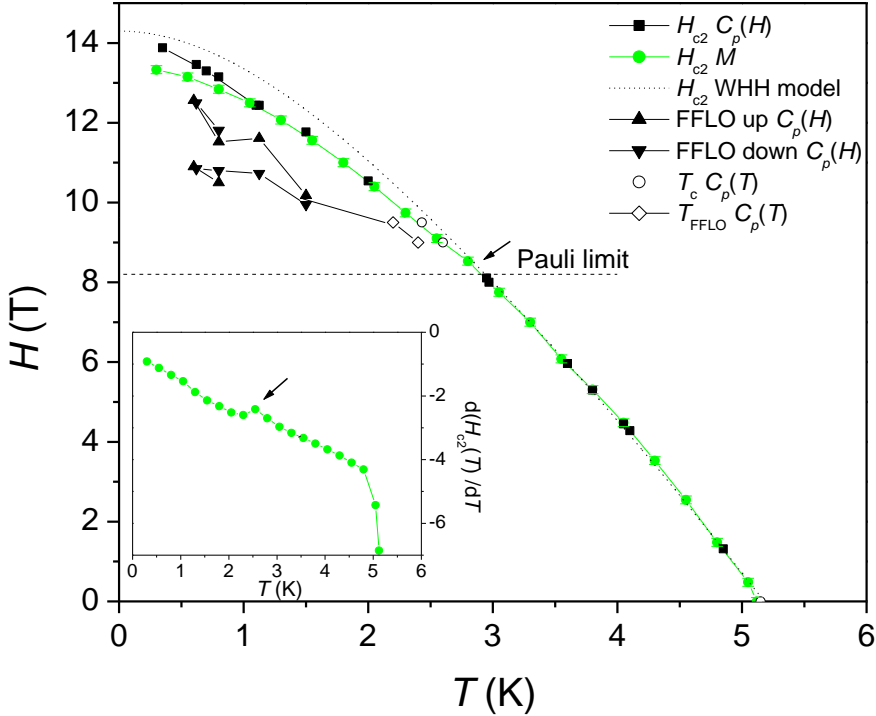

**Fig. 4.** Magnetic field vs. temperature phase diagram of $Ti_4Ir_2O$ compiled from the data shown in Fig. 1 – 3. The horizontal dotted line marks the Pauli limiting field at 8.2 T. The dashed line is a fit with the WHH model for the upper critical field ($H_{c2}$) line of a superconductor. A deviation of the WHH line develops abruptly at 8 T near the Pauli limit (as indicated by the arrow), which is attributed to the Zeeman pairbreaking effect. This is further illustrated in the inset which shows the temperature derivative of the $H_{c2}$ line as determined from the magnetization (green squares), which shows a jump like anomaly at this field. However, the $H_{c2}$ line increases well beyond the Pauli limit with additional phase transition anomalies attributed to transition into a high-field FFLO state. They appear within the superconducting phase with a clear hysteresis. Upright triangles mark anomalies appearing during increasing field in different quantities, while downright triangles are from data measured in decreasing field.

As the maximum of the magnetization peaks approximately coincide with the specific heat FFLO transitions measured in increasing field, it is likely an enhanced pinning related to the spatial modulation of the order parameter in the FFLO phase which causes the magnetization peak effect here. The alternating phase modulation related to the order parameter modulation was predicted to leave distinctive features in the vortex dynamics [28,29], which may be responsible for the magnetization peak in the FFLO region above the Pauli limit.

Our data exhibits characteristic features in the magnetic field vs. temperature phase diagram near and above the Pauli limit for superconductivity, which align with the formation of an FFLO state. The remaining question is why the upper critical field line in this isotropic superconductor can approach the Pauli limit, necessitating a particularly high orbital limit for superconductivity that well exceeds the Pauli limit. In layered superconductors, the absence of closed orbits around the two-dimensional Fermi cylinders when the field is applied parallel to the layered structure either excludes the orbital limit or pushes it up to very high values. While this is not possible in the isotropic superconductor $Ti_4Ir_2O$, its composition of particularly heavy atoms would naively suggest strong spin-orbit coupling as the reason for its high orbital limit, as already discussed Ref. [26]. Other superconductors within the same structural group are known to exhibit exceptionally high upper critical fields. For instance, $Ti_4Ru_2O$ approaches the Pauli limit without significantly exceeding it, while $Ti_4Co_2O$ demonstrates a violation of the Pauli limit, albeit less than $Ti_4Ir_2O$ [25]. Co and Ir have a substantial difference in spin-orbit coupling, but only a small difference in the degree of violation of the Pauli limit is observed in $Ti_4Co_2O$ and $Ti_4Ir_2O$, suggesting that the spin-orbit coupling cannot be the main factor determining the high orbital limit. The cause of the high orbital limits in this class of materials remains elusive. It may be attributed to the intricate nature of their unit cells and band structure. It has been proposed that $Ti_4Ir_2O$ may possibly be a multiband superconductor with multiple superconducting gaps [26], which is known to increase the orbital limit for superconductivity, as demonstrated in the prominent case of $MgB_2$ [30]. Bandstructure calculations indeed reveal the presence of multiple bands crossing the Fermi level [25], but this requires further investigation beyond the scope of this work.

A peculiar feature is observed in the specific heat data set taken at 0.6 and 0.8 K. While they show very similar small step-like FFLO anomaly as the data taken at higher temperatures, an additional phase transition anomaly occurs at ~10.8 T. This anomaly is reproducible and occurs at approximately the same applied field upon increasing or decreasing the field. Multiple phase transitions within the FFLO phase with multiple **q** phases have indeed been predicted [31-33], and this may be an example of such a segmentation of the FFLO phase.

In conclusion, using thermodynamic bulk probes, we have reported characteristic features in the magnetic field vs. temperature phase diagram of the isotropic superconductor $Ti_4Ir_2O$. These features align with the possible formation of an FFLO state and thus provide a plausible explanation for the violation of the Pauli limit in this superconductor. This adds a fully isotropic superconductor to the list of potential FFLO superconductors, providing new insights for contemporary theories of the FFLO state.

# Methods

### Sample Preparation

The preparation of the polycrystalline $Ti_4Ir_2O$ samples by solid state reactions and their characterization is described in detail in Ref. [26]. The sample under investigation represented a millimeter sized conglomerate of crystallites of sized of average size in the range of 1 – 10 μm. These individual crystallites are large enough to be considered as bulk samples, but our thermodynamic probes required larger masses. The high quality can be judged from the relatively sharp zero field specific heat transition (Fig. 2), which is sharper than what was reported for single crystalline samples [25].

### Experimental techniques

The magnetization of the sample was measured using a Faraday balance. The sample was affixed to a circular plate at the end of a cantilever, which also serves as one of the plates of a parallel plate capacitor. The balance is strategically positioned slightly above the center of a 15T superconducting magnet within a $^3$He cryostat, resulting in a marginally inhomogeneous field. This field induces a force on the diamagnetic isotropic superconducting sample, pushing it against the cantilever and altering the sensor's capacitance. This change is measured using a General Radio Capacitance bridge, coupled with a digital SR850 lock-in amplifier.

The specific heat of the sample was determined using a custom-made AC calorimeter [19,21]. This calorimeter consists of a sapphire chip with a resistive heater and thermometer. The chip is suspended on thin phosphor bronze wires, which provide electrical contacts and a thermal link to the thermal bath formed by the $^3$He probe. Given that the FFLO transition at low temperatures typically occurs at a relatively constant field, most specific heat data was measured by periodically modulating the temperature of the chip and sample with an amplitude of ~1mK during a field sweep at a rate of 0.1 T/min [19,21], while carefully stabilizing the average chip temperature. This method probes the sample's response to minor temperature variations and provides a specific heat curve as a function of the applied field at a constant average sample temperature. Temperature dependent measurements in constant applied magnetic fields have been performed by sweeping the temperature of the thermal bath at a rate of +0.1 K/min. A DC current of the order of 100 nA is sent through the thermometer, while the AC output voltage of the lock-in amplifier is sent to the resistive heater to cause the temperature modulation. The phase and amplitude of the alternating voltage across the thermometer are then measured by the lock-in amplifier.

# REFERENCES

[1] Fulde, P, Ferrell, R. A. (1964). Superconductivity in a strong spin-exchange field. Phys Rev 135 A550. https://doi.org/10.1103/PhysRev.135.A550.
[2] Larkin, A. I., Ovchinnikov Yu. N. (1965). Nonuniform state of superconductors. Sov. Phys. JETP 20 762.
[3] Clogston, A. K. (1962). Upper Limit for the Critical Field in Hard Superconductors. Phys. Rev. Lett. 9 266. https://link.aps.org/doi/10.1103/PhysRevLett.9.266.
[4] Chandrasekhar, B. S. (1962). A note on the maximum critical field of high-field superconductors, Appl. Phys. Lett. 1 7. https://doi.org/10.1063/1.1777362.
[5] Radovan, H. A., Fortune, N. A., Murphy, T. P., Hannahs, S. T., Palm, E. C., Tozer, S. W., Hall, D. (2003). Magnetic enhancement of superconductivity from electron spin domains Nature (London) 425 51. https://doi.org/10.1038/nature01842.
[6] Bianchi A, Movshovich R, Capan C, Pagliuso P G, Sarrao J L 2003 Possible Fulde-Ferrell-Larkin-Ovchinnikov Superconducting State in CeCoIn$_5$ Phys. Rev. Lett. 91 187004. https://link.aps.org/doi/10.1103/PhysRevLett.91.187004.
[7] Kenzelman, M., Gerber, S., Egetenmeyer, N., Gavilano, J. L., Strässle, Th., Bianchi, A. D., Ressouche, E., Movshovich, R., Bauer, E. D., Sarrao, J. L., Thompson, J. D. (2010). Evidence for a Magnetically Driven Superconducting Q Phase of CeCoIn$_5$ Phys. Rev. Lett. 104 127001. https://link.aps.org/doi/10.1103/PhysRevLett.104.127001.
[8] Singleton, J., Symington, J. A., Nam, M. S., Ardavan, A., Kurmoo, M., Day, P. (2000). Observation of the Fulde-Ferrell-Larkin-Ovchinnikov state in the quasi-two-dimensional organic superconductor κ-(BEDT-TTF)$_2$Cu(NCS)$_2$ J. Phys. Condens. Matter 12 L641. https://doi.org/10.1088/0953-8984/12/40/102.
[9] Lortz, R., Wang, Y., Demuer, A., Böttger, P. H. M., Bergk, B., Zwicknagl, G., Nakazawa, Y., Wosnitza, J. (2007). Calorimetric Evidence for a Fulde-Ferrell-Larkin-Ovchinnikov Superconducting State in the Layered Organic Superconductor κ-(BEDT-TTF)$_2$Cu(NCS)$_2$ Phys. Rev. Lett. 99 187002. https://link.aps.org/doi/10.1103/PhysRevLett.99.187002.
[10] Bergk, B., Demuer, A., Sheikin, I., Wang,Y., Wosnitza, J., Nakazawa, Y., Lortz, R. (2011). Magnetic torque evidence for the Fulde-Ferrell-Larkin-Ovchinnikov state in the layered organic superconductor κ-(BEDT-TTF)$_2$Cu(NCS)$_2$ Phys. Rev. B 83 064506. https://link.aps.org/doi/10.1103/PhysRevB.83.064506.
[11] Beyer, R., Bergk, B., Yasin, S., Schlueter, J. A., Wosnitza, J. (2012). Angle-Dependent

Evolution of the Fulde-Ferrell-Larkin-Ovchinnikov State in an Organic Superconductor Phys. Rev. Lett. 109 027003. https://link.aps.org/doi/10.1103/PhysRevLett.109.027003.
[12] Koutroulakis, G., Kühne, H., Schlueter, J. A., Wosnitza, J., Brown, S. E. (2016). Microscopic Study of the Fulde-Ferrell-Larkin-Ovchinnikov State in an All-Organic Superconductor Phys. Rev. Lett. 116 067003. https://link.aps.org/doi/10.1103/PhysRevLett.116.067003.
[13] Tsuchiya, S., Yamada, J.-i., Sugii, K., Graf, D., Brooks, J. S., Terashima, T., Uji, S. (2015). Phase boundary in a superconducting state of κ-(BEDT-TTF)$_2$Cu(NCS)$_2$: Evidence of the Fulde–Ferrell–Larkin–Ovchinnikov phase J. Phys. Soc. Jpn. 84 034703. https://doi.org/10.7566/JPSJ.84.034703.
[14] Agosta, C. C., Jin, J., Coniglio, W. A., Smith, B. E., Cho, K., Stroe, I., Martin, C., Tozer, S. W., Murphy, T. P., Palm, E. C., Schlueter, J. A., Kurmoo, M. (2012). Experimental and semiempirical method to determine the Pauli-limiting field in quasi two-dimensional superconductors as applied to κ-(BEDT-TTF)$_2$Cu(NCS)$_2$: Strong evidence of a FFLO state Phys. Rev. B 85 214514. https://doi.org/10.1103/PhysRevB.85.214514.
[15] Wright, J. A., Green, E., Kuhns, P., Reyes, A., Brooks, J., Schlueter, J., Kato, R., Yamamoto, H., Kobayashi, M., Brown, S. E. (2011). Zeeman-Driven Phase Transition within the Superconducting State of κ-(BEDT-TTF)$_2$Cu(NCS)$_2$ Phys. Rev. Lett. 107 087002. https://doi.org/10.1103/PhysRevLett.107.087002.
[16] Mayaffre, H., Kramer, S., Horvatić, M., Berthier, C., Miyagawa, K., Kanoda, K., Mitrović, V. F. (2014). Evidence of Andreev bound states as a hallmark of the FFLO phase in κ-(BEDT-TTF)$_2$Cu(NCS)$_2$ Nat. Phys. 10 928. https://doi.org/10.1038/nphys3121.
[17] Coniglio, W. A., Winter, L. E., Cho, K., Agosta, C. C., Fravel, B., Montgomery, L. K. (2011). Superconducting phase diagram and FFLO signature in λ-(BETS)$_2$GaCl$_4$ from rf penetration depth measurements, Phys. Rev. B 83 224507. https://doi.org/10.1103/PhysRevB.83.224507.
[18] Cho, K., Smith, B. E., Coniglio, W. A., Winter, L. E., Agosta, C. C., Schlueter, J. A. (2009). Upper critical field in the organic superconductor β''-(ET)$_2$SF$_5$CH$_2$CF$_2$SO$_3$: Possibility of Fulde-Ferrell-Larkin-Ovchinnikov state Phys. Rev. B 79 220507(R). https://doi.org/10.1103/PhysRevB.79.220507.
[19] Cho, C.-w., Yang, J. H., Yuan, N. F. Q., Shen, J., Wolf, T., Lortz, R. (2017) Thermodynamic Evidence for the Fulde-Ferrell-Larkin-Ovchinnikov State in the KFe$_2$As$_2$ Superconductor Phys. Rev. Lett. 119 217002. https://doi.org/10.1103/PhysRevLett.119.217002.
[20] Kasahara, S., Sato, Y., Licciardello, S., Čulo, M., Arsenijević, S., Ottenbros, T., Tominaga, T., Böker, J., Eremin, I., Shibauchi, T., Wosnitza, J., Hussey, N. E., Matsuda, Y. (2020) Evidence for an Fulde-Ferrell-Larkin-Ovchinnikov State with Segmented Vortices in the BCS-BEC-Crossover Superconductor FeSe Phys. Rev. Lett. 124 107001. https://doi.org/10.1103/PhysRevLett.124.107001.
[21] Cho, C.-w., Lyu, J., Ng, C. Y., He, J. J., Lo, K. T., Chareev, D., Abdel-Baset, T. A., Abdel-Hafiez, M., Lortz, R. (2021). Evidence for the Fulde–Ferrell–Larkin–Ovchinnikov state in bulk NbS$_2$, Nat Commun 12, 3676. https://doi.org/10.1038/s41467-021-23976-2.
[22] Wan, P., Zheliuk, O., Yuan, N. F. Q., Peng, X., Zhang, L., Liang, M., Zeitler, U., Wiedmann, S., Hussey, N. E., M. Palstra, T. T., Ye, J. (2023). Orbital Fulde–Ferrell–Larkin–Ovchinnikov state in an Ising superconductor, Nature 619, 46-51. https://doi.org/10.1103/PhysRevResearch.5.043122.
[23] Gor'kov L. P. (1960), The critical supercooling field in superconductivity theory Sov. Phys. JETP 10 593.
[24] Lee, P. A. (2014). Amperean Pairing and the Pseudogap Phase of Cuprate Superconductors Phys. Rev. X 4 031017. https://doi.org/10.1103/PhysRevX.4.031017.

[25] Ma, K. Y., Lefèvre, R., Gornicka, K., Jeschke, H. O., Zhang, X., Guguchia, Z., Klimczuk, T., von Rohr, F. O. (2021). Group-9 Transition-Metal Suboxides Adopting the Filled-Ti$_2$Ni Structure: A Class of Superconductors Exhibiting Exceptionally High Upper Critical Fields, Chem. Mater. 33, 8722−8732. https://link.aps.org/doi/10.1103/PhysRevB.110.174507.

[26] Ruan, B.-B., Zhou, M.-H., Yang, Q.-S., Gu, Y.-D., Ma, M.-W., Chen, G.-F., Ren, Z.-A. (2022). Superconductivity with a Violation of Pauli Limit and Evidences for Multigap in $\eta$-Carbide Type Ti$_4$Ir$_2$O, Chin. Phys. Lett. 39, 027401. https://doi.org/10.1088/0256-307X/39/2/027401.

[27] Lortz, R., Musolino, N., Wang, Y., Junod, A., Toyota, N. (2007). Origin of the magnetization peak effect in the Nb$_3$Sn superconductor, Phys. Rev. B 75, 094503. https://doi.org/10.1103/PhysRevB.75.094503.

[28] Bulaevskii, L., Buzdin, A. & Maley, M. (2003). Intrinsic pinning of vortices as a direct probe of the nonuniform Larkin-Ovchinnikov-Fulde-Ferrell state in layered superconductors. Phys. Rev. Lett. 90, 067003. https://doi.org/10.1103/PhysRevLett.90.067003.

[29] Buzdin, A., Matsuda, Y., Shibauchi (2007). T. FFLO state in thin superconducting films. Europhys. Lett. 80, 67004. https://doi.org/10.1209/0295-5075/80/67004.

[30] Wang, Y., Plackowski, T., Junod, A. (2001). Specific heat in the superconducting and normal state (2-300 K, 0-16 Teslas), and magnetic susceptibility of the 38-K superconductor MgB$_2$: evidence for a multicomponent gap, Physica C 355, 179-193. https://doi.org/10.1016/S0921-4534%2801%2900617-7.

[31] Shimahara, H. (1998). Structure of the Fulde-Ferrell-Larkin-Ovchinnikov state in two-dimensional superconductors. J. Phys. Soc. Jpn 67, 736–739. https://doi.org/10.1143/JPSJ.67.736.

[32] Bowers, J. A., Rajagopal, K. (2002). Crystallography of color superconductivity. Phys. Rev. D 66, 065002. https://doi.org/10.1103/PhysRevD.66.065002.

[33] Matsuda, Y. & Shimahara, H. (2007). Fulde–Ferrell–Larkin–Ovchinnikov state in heavy fermion superconductors. J. Phys. Soc. Jpn 76, 051005. https://doi.org/10.1143/JPSJ.76.051005.

# Data availability

The experimental data supporting the findings of this work are available from the corresponding author upon reasonable request.

# Acknowledgements

We thank U. Lampe for technical support. This work was supported by grants from the Research Grants Council of the Hong Kong Special Administrative Region, China (GRF-16303820, C6025-19G-A) and the National Natural Science Foundation of China (Grant No. 12074414).

## Author Contributions

This work was initiated by R.L., J.H. carried out the experiments with help from Y.H.N. and O.A. The Ti$_4$Ir$_2$O sample was provided by B.B.R and Z.A.R. The manuscript was prepared by R.L. and all authors were involved in discussions and contributed to the manuscript.

## Competing interests

The authors declare no competing interests.